\documentclass[%
reprint,
superscriptaddress,
groupedaddress,
amsmath,amssymb,
aps,
prb,
]{revtex4-2}

\usepackage{graphicx}
\usepackage{dcolumn}
\usepackage{bm}
\usepackage{color}
\usepackage{ulem}

\newcommand{\ETNCS}{$\kappa$-(BEDT-TTF)$_2$Cu(NCS)$_2$}
\newcommand{\NCS}{$\kappa$-NCS}

\begin{document}

\title{A sturdy spin-momentum locking in a chiral organic superconductor}

\author{Takuro Sato}
\email{Corresponding author : takurosato@ims.ac.jp}
\affiliation{Research Center of Integrative Molecular Systems (CIMoS), Institute for Molecular Science, Myodaiji, Okazaki, Aichi, 444-8585, Japan}
\affiliation{the Graduate University for Advanced Studies, Myodaiji, Okazaki, Aichi, 444-8585, Japan}

\author{Hiroshi Goto}
\affiliation{Research Center of Integrative Molecular Systems (CIMoS), Institute for Molecular Science, Myodaiji, Okazaki, Aichi, 444-8585, Japan}
\affiliation{the Graduate University for Advanced Studies, Myodaiji, Okazaki, Aichi, 444-8585, Japan}

\author{Hiroshi M. Yamamoto}
\email{Corresponding author : yhiroshi@ims.ac.jp}
\affiliation{Research Center of Integrative Molecular Systems (CIMoS), Institute for Molecular Science, Myodaiji, Okazaki, Aichi, 444-8585, Japan}
\affiliation{the Graduate University for Advanced Studies, Myodaiji, Okazaki, Aichi, 444-8585, Japan}
\affiliation{Quantum Research Center for Chirality (QuaRC), Institute for Molecular Science, Myodaiji, Okazaki, Aichi 444-8585, Japan}


\begin{abstract}
Among noncentrosymmetric structures, chirality has recently been recognized as a novel source of asymmetrical charge/spin transports as exemplified by electrical magnetochiral anisotropy (EMChA) and chirality-induced spin selectivity. 
Although similar bulk-charge rectification and Rashba-Edelstein effect in polar systems are quantitively reproducible by theory based on the electronic band structures, the relevance of band parameters in chiral effects remains elusive. 
Here, by working with a chiral organic superconductor, we experimentally demonstrate a gigantic EMChA and large superconducting diode effect, both of which are difficult to be explained solely by its band parameters. 
A two-critical-current signature and an enhanced critical field suggested triplet-mixed Cooper pairs with anomalously enhanced spin-orbit coupling above atomic limit. 
Our results clearly highlight a unique spin-momentum locking with large stiffness beyond the expectation, suggesting an unknown driving force for spin polarization inherent to chirality.

\end{abstract}

\maketitle


\section{\label{sec:level1}INTRODUCTION}

Skewed electron's motion in non-centrosymmetric materials is linked to the direction of its spin through asymmetric spin-orbit coupling (SOC), so that the spin can no longer rotate freely\cite{dresselhaus1955, rashba1960, bychkov1984}. 
This phenomenon, known as spin-momentum locking, has been primarily investigated in polar systems\cite{ast2007}. 
One of its consequences is current-induced magnetization (Edelstein effect), in which polarized spins whose direction is perpendicular to the electron flow emerge\cite{edelstein1990, ganichev2002, nomura2015, nakayama2016}.
Additionally, such spin-momentum locked electric band under magnetic field gives rise to a unique bulk-charge-rectifying effect\cite{ideue2017, yasuda2016, he2019a}. 
This concept has been further extended to non-centrosymmetric superconductivity, yielding superconducting diode effect (SDE), where the superfluid flow can be switched on and off depending on the direction of the current\cite{edelstein1996, ando2020}. 
These phenomena are quantitatively described in terms of their electronic band parameters with nominal SOC.

Among non-centrosymmetric materials, chiral system that possesses no roto-reflection symmetry has been a focus of renewed interest since the discovery of chirality-induced spin selectivity (CISS) in chiral organic molecules\cite{naaman2012, naaman2019, ray1999, gohler2011}. 
CISS is a spin-rectifying effect consistent with a chiral version of current-induced magnetization, that is, electron spin aligns parallel or antiparallel to the electron velocity, depending on the handedness of the material. 
Remarkably, a significant spin-rectification efficiency observed even at room temperature in organics, despite negligible SOC, underscores the peculiarity of chirality. 
In addition to the CISS effect, the bulk-charge-rectifying effect in chiral materials called electrical magnetochiral anisotropy (EMChA) appears insensitive to SOC strength\cite{rikken2001, pop2014, krstic2002, rikken2019, yokouchi2017, aoki2019}. 
The unexpectedly large efficiencies observed in these two chirality-driven spin/charge rectifications have sparked an intriguing inquiry into the presence of a nontrivial spin-charge coupling inherent to chirality, unlike polar systems. 
This issue, however, remains a fundamental open question due to the lack of a suitable experimental platform for quantitatively comparing the rectification efficiency between chiral and polar symmetries.

\begin{figure*}
\includegraphics{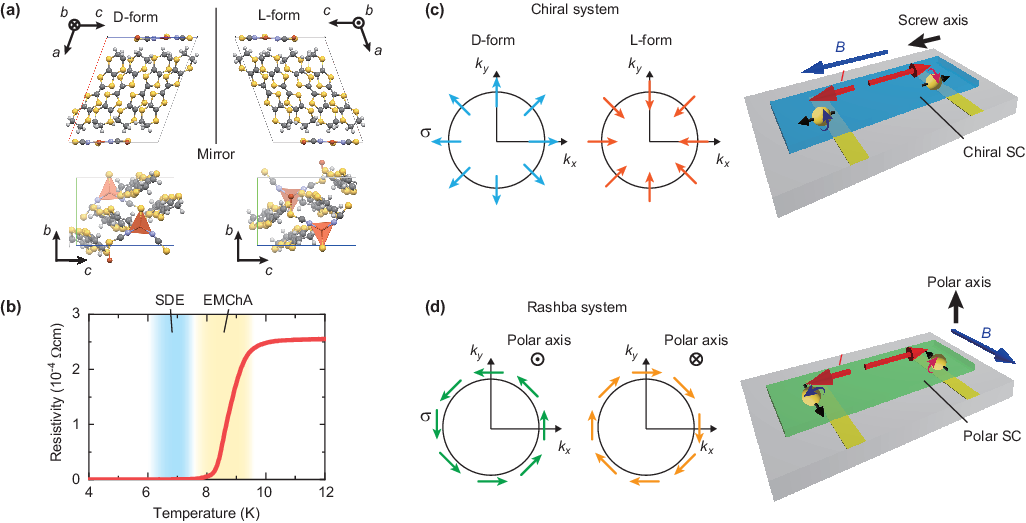}
\caption{
Basic properties of a chiral organic superconductor and schematics of charge- and spin-rectification effect arising from several spin-momentum locking bands.
(a) Crystal structures of chiral organic superconductor, {\ETNCS}, for D- and L-forms. 
Neither the conducting BEDT-TTF nor the insulative anion layers themselves are chiral; the former possesses two-fold rotational symmetry along the $b$-axis due to oblique arrangement of BEDT-TTF molecules, while the latter has polar symmetry along the same $b$-axis (represented by red triangles).
As a consequence, a three-dimensional arrangement of these two layers defines a ${2_1}$-screw axis pointing to the in-plane $b$-axis, resulting in chiral structure. 
(b) Temperature dependence of the resistivity of our thin-film {\NCS} around $T_{\rm{c}}$ under zero magnetic field. 
(c, d) Chiral- and Polar-type spin-momentum locking and expected charge- and spin-rectification in the experimental configuration, respectively. 
$k_{\rm{x}}$, $k_{\rm{y}}$, $\sigma$, $B$, and $I$ denote wave vectors along $x$ and $y$ direction in 2D plane, electric spin, magnetic field, and electric current, respectively.
}
\label{Fig1} 
\end{figure*}

The chiral organic superconductor, {\ETNCS} (hereafter {\NCS}), offers a unique opportunity to this issue thanks to the following advantages: (i) a giant CISS effect has been proven in  {\NCS}, in which resultant spin polarization is far beyond the expected one with organic SOC\cite{nakajima2023}, (ii) solid microscopic theories for the charge-rectification effect including EMChA in non-centrosymmetric superconductivity are available\cite{he2020, wakatsuki2018, hoshino2018}, and (iii) the charge-rectification effects in various polar-type superconductors have been experimentally investigated. 
Therefore, it is crucial to examine whether the chiral organic superconductor can exhibit EMChA beyond its SOC, and to compare it with polar systems. 
Herein, we report such study with the chiral superconductor by demonstrating gigantic EMChA, large chiral superconducting diode effect, and two-critical currents in superfluid states, all of which point to emergence of strongly spin-momentum-locked Cooper pairs exhibiting a triplet-mixed state with effectively enhanced SOC. 
Although the type of spin carrier was unknown in our previous CISS experiments, our present results allow us to discuss chiral-type spin-momentum locking, namely parallel alignment of momentum and spin ($\bm{k} \cdot \bm{\sigma}$) as shown in Fig.~1(c), of superfluid responsible for the giant CISS effect as well.

\section{\label{sec:level1}BASIC PROPERTIES OF {\NCS} AND DEVICE PREPARATIONS}

{\NCS} is comprised of conducting BEDT-TTF layers and insulating anion layers, wherein BEDT-TTF denotes bis(ethylenedithio)tetrathiafulvalene [Fig.~1(a)]. 
In the conducting layers, one hole per a pair of BEDT-TTFs is compensated by counter anion, resulting in an effectively-half-filled band\cite{kino1996}. 
Importantly, {\NCS} belongs to a space group of ${P2_1}$, which allows its crystal form to be chiral with ${2_1}$-screw axis pointing to the in-plane $b$-axis [Fig.~1(a)]\cite{fujio2009, urayama1988}. 
This situation is not the case for similar $\kappa$-type compounds, $\kappa$-(BEDT-TTF)$_2$Cu[N(CN)]$_2$Br or $\kappa$-(BEDT-TTF)$_2$Cu[N(CN)]$_2$Cl, whose space group is $P$$_{\rm{nma}}$ that is not chiral.
For thre present study, we prepared a thin-film single crystals of {\NCS} on the plastic substrates, which allows us to obtain a homo-chiral domain and to reach high current density (the details of the device preparation is shown in Supplemental Materials\cite{SM}).
A well-defined superconductivity sets in at $T_{\rm{c}}$ of 8.7~K (midpoint) in our device [Fig.~1(b)]. 
Note that spin-orbit-coupled charge character has been disregarded in vast majority of past researches on the system, since magnitude of a SOC in $\kappa$-type BEDT-TTF is estimated to be no more than 1-2~meV\cite{winter2017} that is roughly equivalent to the order of $10^{-2}$~eV$\cdot \rm{\AA}$ (see Supplemental Materials\cite{SM}).

\section{\label{sec:level1}RESULTS}

\subsubsection{Gigantic electrical magnetochiral anisotropy in a chiral superconductor}

\begin{figure*}
\includegraphics{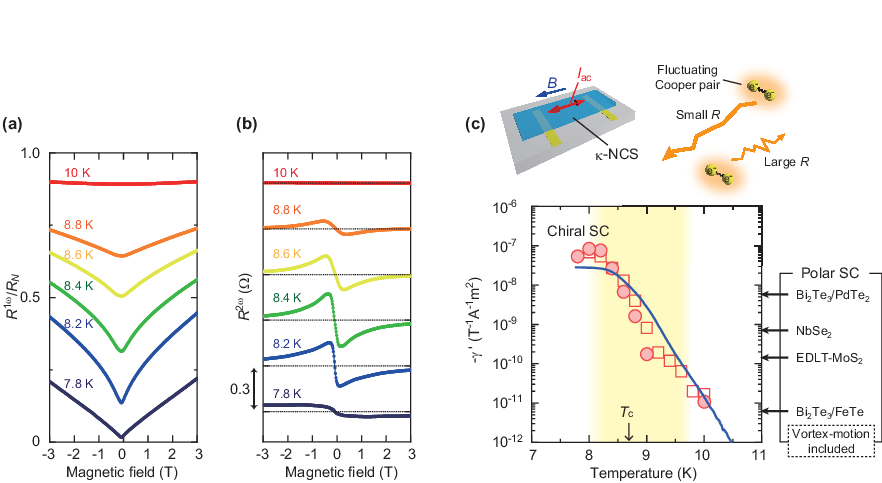}
\caption{
Giant EMChA in superconducting fluctuation regime. 
(a, b) First and second harmonic magnetoresistance $R^{1\omega}$, and $R^{2\omega}$, respectively) across $T_{\rm{c}}$ (= 8.7~K) as a function of in-plane magnetic field $\bm{B}$ parallel to $\bm{I}$ ($j_{\rm{ac}}$=5.2$\times$10$^6$ A/m$^2$). 
The dataset was measured while decreasing the magnetic field. 
(c) Temperature dependence of the normalized nonreciprocal parameter, ${\gamma'}$, evaluated from zero field resistance in (a) and the slope around zero field in (b). 
Blue line shows the fitting curve of  ${\gamma'}(T)={\gamma}_{\rm{s}} S \left\{1-\frac{R(T)}{R_{\rm{N}}} \right\}^2$.
The fitting is performed for the yellow shaded region where fluctuation of the Copper pair is relevant for producing finite resistance (the schematics is shown in the inset). 
Normal-state resistance $R_{\rm{N}}$ is defined at $T$ = 11~K. 
The plot of red circles and open square represent dataset obtained at different position of the same device, confirming that the device has a single chiral domain. 
To evaluate the order of magnitude of ${\gamma'}$ values for other 2D or interfacial polar-type superconductors, we assume that the effective thickness of the superconductivity is 1~nm, which roughly corresponds to the monolayer thickness of transition metal dichalcogenides. 
It is reasonable to expect that the effective thickness of the superconducting region in gate-induced or interfacial superconductivity is comparable to the monolayer thickness. 
The estimated of ${\gamma'}$ values in these polar systems are shown on the right $y$-axis.
}
\label{Fig2} 
\end{figure*}

We first test whether EMChA emerges in superconducting phase of {\NCS} with an observable level, based on the established equation,  $R(\bm{I},\bm{B}) = R_0\left(1+{\gamma}\bm{I}{\cdot}\bm{B} \right)$, where $R$, $\bm{I}$ and $\bm{B}$ are resistance, current and magnetic field, respectively. 
A chiral parameter $\gamma$ gives rise to a nonreciprocal resistance whose sign relies on the handedness of the material\cite{rikken2001}. 
Unlike the required geometry for bulk rectification in polar system ($\bm{I} \perp \bm{B}$), our measurements have been carried out for both electric current and magnetic field directed in parallel ($\bm{I} \parallel \bm{B}$) along the screw axis to investigate the chiral response in {\NCS} [see Figs.~1(c) and 1(d)]. 
Since the $\gamma$ term appears in measured voltage as a nonlinear response which is proportional to $I^2$, we measured second harmonic resistance, $R^{2\omega}$, under the application of a.c.~excitation, $j_{\rm{ac}}$.
Figures~2(a) and 2(b) show the magnetic field dependence of first harmonic resistance, $R^{1\omega}$, and $R^{2\omega}$ at representative temperatures across $T_{\rm{c}}$, respectively.
Positive magnetoresistance in $R^{1\omega}$ reflects that superconductivity is gradually suppressed by magnetic field [Fig.~2(a)]. 
On approaching $T_{\rm{c}}$ from 10~K, the asymmetric peaks in $B$--$R^{2\omega}$ profile are markedly developed [Fig.~2(b)]. 
Simultaneously,  $R^{2\omega}$ around 0~T exhibits a linear dependence on the magnetic field above 7.8~K.
Both observations signal the appearance of the EMChA in superconducting regime.
Additionally, we show that the peak intensity varies as a function of the in-plane magnetic field angle against the $b$-axis, roughly following a cosine curve (Fig.~S5 in the Supplemental Materials\cite{SM}). 
This behavior is consistent with the expected characteristics based on chiral symmetry, further confirming that the observed asymmetric peaks originate from EMChA.
We note that in the experimental setup with $\bm{I} \parallel \bm{V}  \parallel \bm{B}$, the field-induced vortices aligned with in-plane current direction are immobile, and also cannot generate lateral resistance by possible 2$\omega$ Joule heating, ruling out the thermoelectric effects such as Nernst effect as a factor in our asymmetric $R^{2\omega}$ signals. 
Furthermore, it is unlikely that inhomogeneous supercurrent paths around $T_{\rm{c}}$ are responsible for the observed $R^{2\omega}$ profile. 
This is because any extrinsic chirality, arising from broken mirror symmetries due to the local sample inhomogeneities, would be random, which is inconsistent with both the observed angular dependence and the fixed sign of the asymmetric $R^{2\omega}$ as the magnetic field and electric current are varied (Figs.~S5 and S7 in the Supplemental Materials\cite{SM}).

The temperature profile of the normalized nonreciprocal coefficient, $\gamma'=\gamma S=\frac{2R^{2\omega}}{R^{1\omega}BI} S $, in which $S$ is a cross-sectional area, is evaluated based on the slope of $B$--$R^{2\omega}$ profile around 0~T [Fig.~2(c)].
The ${\gamma'}$ suddenly increases in the superconducting state by more than three orders of magnitude. 
Despite its small SOC, surprisingly, the observed ${\gamma'}$ in {\NCS} well surpasses previous records in inorganic polar superconductors with heavy elements [Fig.~2(c)]\cite{wakatsuki2017, yasuda2019, zhang2020, masuko2022}.
It is noteworthy that the dominant factor for EMChA in our chiral superconductor is solely superconducting fluctuations as vortices remain immobile in the $\bm{I} \parallel \bm{B}$ geometry. 
In contrast, for polar superconductors in the $\bm{I} \perp \bm{B}$ geometry, the nonreciprocal signal is known to arise from both superconducting fluctuations and vortex motion, with the latter usually being the larger contribution\cite{itahashi2020}.
Given this difference, larger ${\gamma'}$ observed in {\NCS} compared to that in polar superconductors is significant; the actual contribution only from superconducting fluctuations in the polar case is likely much smaller than the level indicated in Fig.~2(c). 
In terms of differences between chiral and polar systems, we also point that the appearance of EMChA in {\NCS} is limited in the much narrower field range around 0~T compared with those in polar superconductors. 
We attribute this to the peculiar coupling between superconductivity and structural chirality in {\NCS}, as discussed in depth in ``Discussion about a coupling between superconductivity and structural chirality in {\NCS}'' section in Supplemental Materials\cite{SM}.

A microscopic insight into the origin of the giant EMChA in the chiral superconductivity can be provided by the recent theory that relies on the prevalent notion that Cooper pairs in non-centrosymmetric superconductors are no longer purely spin-singlet type, but of mixed spin-triplet and spin-singlet type (namely, parity mixing)\cite{wakatsuki2018, hoshino2018}. 
This theory proposes that the EMChA in such a superconductor is due to superconducting fluctuations with the mixed singlet and triplet pairings, and the enhancement of the ${\gamma'}$ towards $T_{\rm{c}}$ is predicted to follow ${\gamma'}(T)={\gamma}_{\rm{s}} S \left\{1-\frac{R(T)}{R_{\rm{N}}} \right\}^2$, where $\gamma_s$ is a temperature-independent parameter and $R_{\rm{N}}$ is a resistance of normal state (the expression of ${\gamma'}$ is derived in \cite{hoshino2018}, and is used to analyze the experimental data for polar superconductor in \cite{itahashi2020}). 
As shown in Fig.~2(c), the predicted ${\gamma'}(T)$ with using observed $R(T)$ accurately reproduces the observed temperature dependence of ${\gamma'}$, including the saturated behavior below 8.2~K.
This confirms that the model is applicable to our system. 
The fitting yields a ${\gamma}_{\rm{s}}$ value of 4500~T$^{-1}$A$^{-1}$, which can be used to assess the degree of the parity mixing, $r_{\rm{t}}=\frac{2V^u}{V^g+V^u}$  ($V^g$ and $V^u$ are singlet and triplet pairing interaction, respectively) through the relationship of ${\gamma}_{\rm{s}}=\frac{\pi \mu_{\rm{B}}\hbar S_{3}}{2WeS_{1}k_{\rm{B}}T_{\rm{c}}} \alpha r_{\rm{t}}$ with the fixed strength of SOC, $\alpha$ (see ``Analysis of EMChA in a non-centrosymmetric superconductor with parity-mixing'' section in Supplemental Materials for more detail\cite{SM}). 
We found that an assumed $\alpha$ of 10$^{-2}$ eV$\cdot \rm{\AA}$ based on the nominal SOC in {\NCS} results in an unphysically large $r_{\rm{t}}$ of $\sim$500, exceeding the pure triplet case ($r_{\rm{t}}$=2) by two orders of magnitude. 
This unacceptable outcome of the analysis casts a doubt on the assumption and strongly suggests a significant enhancement in the effective SOC in the chiral superconductor beyond typical organic SOC levels. 
Considering the relevance of the theoretical model with triplet-mixed fluctuations, the enhanced SOC also indicates an emergence of strong spin-momentum locking in the triplet-mixed Cooper pairs\cite{he2020, edelstein1995, edelstein2003, he2019}. 
Although the reason why chirality is so effective in enhancing the SOC is yet to be clarified, our observations confirm that a combination of chirality and superconducting fluctuations plays the decisive role in enhancing SOC.
We can safely conclude that the polarity along the $b$-axis in {\NCS} is unrelated to the enhanced SOC or triplet mixing. 
This is because, in our setup, both the current and magnetic field are parallel to the polar axis, where polarity-induced nonreciprocity is not symmetrically permitted.

\begin{figure*}
\includegraphics{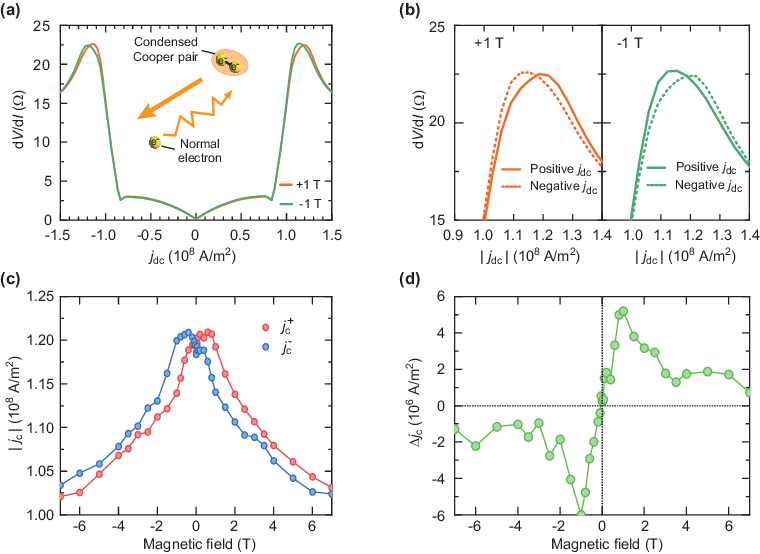}
\caption{
Chiral-version of superconducting diode effect in {\NCS}. 
(a) Differential resistance, $\frac{\rm{d}\it{V}}{\rm{d}\it{I}}$, as a function of d.c.~current bias of the superconducting phase at $T$ =7~K. 
The measurement is performed at magnetic field = +1~T and -1~T (orange and green lines, respectively). 
The inset shows a schematic image of the SDE. 
(b) Comparison in critical current between positive and negative currents at a fixed magnetic field (left : +1~T, right : -1~T). 
(c) Critical current density for positive and negative bias, $j_{\rm{c}}^{+}$ and $|j_{\rm{c}}^{-}|$, and (d) The nonreciprocal component of the critical current $\Delta j_{\rm{c}}$ at 7~K as a function of the magnetic field.
}
\label{Fig3} 
\end{figure*}

\subsubsection{Chiral-type superconducting diode effect}

Since SDE is also a useful tool to elucidate an effective SOC well below $T_{\rm{c}}$, we next pursue the SDE in our chiral organic superconductor by evaluating the critical current of the superconducting phase under the magnetic field. 
In contrast to EMChA, which becomes prominent in the region of superconducting fluctuations near $T_{\rm{c}}$, SDE can emerge at lower temperatures where superconducting fluctuations are frozen. 
Figures~3(a) and 3(b) show a differential resistance, $\rm{d}\it{V}/\rm{d}\it{I}$, as a function of d.c.~bias, $j_{\rm{dc}}$, in superconducting phase at $\pm$1~T. The critical current density, $j_{\rm{c}}$, is defined as $j_{\rm{dc}}$ where the differential resistance forms a peak. 
We detected a discrepancy in $j_{\rm{c}}$ between $\pm$1~T, which is not expected for reciprocal case. 
This unusual behavior is also indicated by a clear difference in $j_{\rm{c}}^{+}$ and $j_{\rm{c}}^{-}$, or the critical current density with a positive and negative $j_{\rm{dc}}$, respectively [Fig.~3(b)]. 
To further visualize the impact of the magnetic ﬁeld, we plotted both $j_{\rm{c}}^{+}$ and $|j_{\rm{c}}^{-}|$, and also the nonreciprocal components of the critical current $\Delta j_{\rm{c}} = j_{\rm{c}}^{+} - |j_{\rm{c}}^{-}|$ as a function of a magnetic field [Figs.~3(c) and 3(d)]. 
Obviously, $\Delta j_{\rm{c}}$ is non-zero at finite magnetic field and moreover its profile is antisymmetric against $B$, in accordance with the expected contribution from $\bm{B} \cdot \bm{I}$. 
With the device$\#$2 exhibiting opposite sign of EMChA compared to those in Fig.~2(a) (device$\#$1), the field profile of $\Delta j_{\rm{c}}$ gives opposite sign (Fig.~S8 in the Supplemental Materials\cite{SM}). 
The one-to-one correspondence between signs of SDE and EMChA ensures that the observed SDE reflects the intrinsic chirality of the bulk superconductivity, demonstrating the chiral version of SDE for the first time.
We found that the amplitudes of both EMChA and SDE in device$\#$2 are several times smaller than those in device$\#$1, which indicates the lower purity of chirality in device$\#$2. 
It is important to emphasize again that our experimental geometry with $\bm{I} \parallel \bm{B}$ cannot induce the vortex motion, which means that any surface barrier for injecting or ejecting vortices, if present, does not dominate the critical current.

We here introduce the diode efficiency, $\eta = 2 \frac{j_{\rm{c}}^{+}-|j_{\rm{c}}^{-}|}{j_{\rm{c}}^{+}+|j_{\rm{c}}^{-}|}$, namely, the $\Delta j_{\rm{c}}$ normalized by the averaged $j_{\rm{c}}$. 
At 7~K, the $\eta$ value reaches approximately 5 \%, which is comparable to that for the SDE in bulk artificial superlattice with heavy elements\cite{ando2020}, again highlighting the anomalously enhanced SOC. 
In Supplemental Materials, we present similar measurements taken at different temperature below $T_{\rm{c}}$, for which amplitude of SDE is not monotonous against temperature and becomes maximum around 7~K (Fig.~S9 in the Supplemental Materials\cite{SM}). 
Although it is premature to quantitatively compare our experimental results with recent theoretical predictions, it is worth noting that the theory on SDE, which assumes spin-momentum-locked normal bands without incorporating spin-triplet mixing, predicts a monotonic enhancement of SDE at lower temperatures\cite{daido2022, nadeem2023}. 
This stands in stark contrast to the non-monotonous temperature dependence observed in our experiments, which may indicate that spin-triplet mixing plays a key role in this discrepancy.

\begin{figure*}
\includegraphics{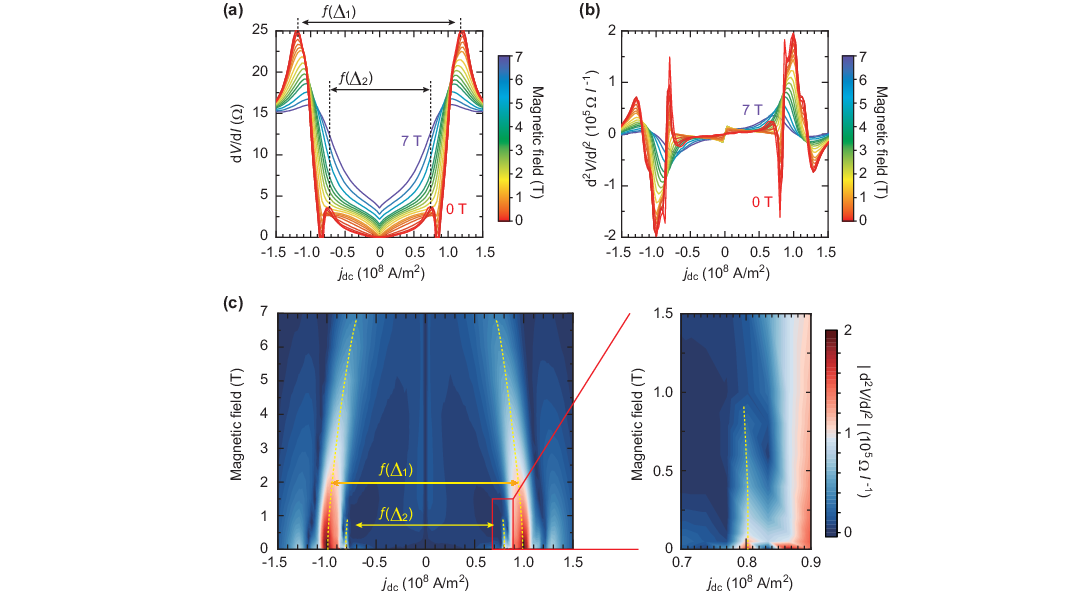}
\caption{
Two-critical-current superconductivity and its magnetic field responses. 
(a, b) Magnetic field dependence of (a) the differential resistance,  $\frac{\rm{d}\it{V}}{\rm{d}\it{I}}$, and (b) its current-derivative, $\frac{\rm{d}^{2}\it{V}}{\rm{d}\it{I}^{\rm{2}}}$, measured at 7~K. 
The color of each plot corresponds to the strength of magnetic field (see color bar). 
(c) The color map of $\left|\frac{\rm{d}^{2}\it{V}}{\rm{d}\it{I}^{\rm{2}}}\right|$ against $j_{\rm{dc}}$ and magnetic field constructed by the dataset in (b) and a close-up view for the sub-critical-current signal around $j_{\rm{dc}}$=0.8$\times$10$^8$ A/m$^2$. 
The yellow dashed lines are the trace of main and sub critical currents. 
The temperature dependence of the differential resistance is displayed in Fig.~S11 in the Supplemental Materials\cite{SM}. 
}
\label{Fig4} 
\end{figure*}

\subsubsection{Two-critical-current superfluid}

In the above experiments, we have assumed mixing of spin-triplet and spin-singlet Cooper pair as a platform for manifestation of enhanced SOC. 
Generally, this spin-singlet and spin-triplet mixing can be accompanied by a formation of two superconducting gaps reflecting the spin-splitting bands. 
Therefore, it is natural to explore the two-gap signature and its response to magnetic field and get further insight into the mixed triplet pairing to justify our previous assumptions. 
Figure~4(a) shows a detailed magnetic field dependence of differential resistance taken at 7~K. 
Without magnetic field, a clear anomaly is identified at an intermediate current regime around 0.8$\times$10$^8$ A/m$^2$ as well as the main peaks of $j_{\rm{c}}$ around 1.2$\times$10$^8$ A/m$^2$.
The observation of two-critical currents ($j_{\rm{c}}$) is consistent with the notion of two-gap superconductivity.
This symptom is also resolved in Fig.~3(a). 
The two-$j_{\rm{c}}$ structure is more visible in the second derivative $\frac{\rm{d}^{2}\it{V}}{\rm{d}\it{I}^{\rm{2}}}$, in which sub $j_{\rm{c}}$ forms a negative peak [Fig.~4(b)]. 
Remarkably, the primary two-$j_{\rm{c}}$ signature is progressively suppressed as magnetic field increases, transforming to the single-$j_{\rm{c}}$ one, which suggests that these two critical currents are distinctive in terms of the response to magnetic field. 
We attempt to track these peaks in more details by constructing $j_{\rm{dc}}$--$B$ map for $|\frac{\rm{d}^{2}\it{V}}{\rm{d}\it{I}^{\rm{2}}}|$ [Fig.~4(c)]; apparently, the main peak in$|\frac{\rm{d}^{2}\it{V}}{\rm{d}\it{I}^{\rm{2}}}|$ persists up to 7~T, whereas the sub-$j_{\rm{c}}$ one immediately vanishes only with 1~T, which implies that pairing mechanisms related to these critical currents are not the same.

A reasonable interpretation for this data posits that the robust critical current more sensitively reflects the mixed triplet character than the sub-critical current does. It is well known that a striking consequence of the triplet is an enhanced $B_{\rm{c2}}^{\parallel}$ above Pauli limit. 
The maximum $B_{\rm{c2}}^{\parallel}$ for singlet superconductivity can be determined solely by Zeeman effect where the Cooper pair is destroyed by flipping one of the spins (called the paramagnetic pair breaking)\cite{chandrasekhar1962}. 
Since triplet pairs are free from the paramagnetic pair breaking, the $B_{\rm{c2}}^{\parallel}$ can be substantially enhanced over Pauli limit. 
To clarify if this is the case in the present system, we examined temperature-dependent resistance for magnetic fields applied perpendicular to and parallel to the conducing plane (Fig.~S2 in the Supplemental Materials\cite{SM}). 
As shown in Fig.~S2 in the Supplemental Materials\cite{SM}, we found that the temperature-evolution of $B_{\rm{c2}}^{\parallel}$ and $B_{\rm{c2}}^{\perp}$, defined as the midpoint of the resistive transition, are well fitted by the phenomenological 2D Ginzburg-Landau (GL) model.
Importantly, the extrapolated $B_{\rm{c2}}^{\parallel}$ based on the GL fitting develops above Pauli limit below 3~K, consistent with the interpretation that the triplet mixing can enhance $B_{\rm{c2}}^{\parallel}$.
Nonetheless, the GL model does not account for the celebrated Fulde–Ferrell–Larkin–Ovchinnikov (FFLO) phase\cite{fulde1964, larkin1964} that also allows $B_{\rm{c2}}^{\parallel}$ to exceed the Pauli limit and has been experimentally identified at very low temperatures in {\NCS}.
Coexistence of the FFLO phase and mixed triplet pairing should be examined in future study.

\section{\label{sec:level1}DISCUSSION}

We have now confirmed that the chiral organic superconductor exhibits both giant EMChA and giant CISS, allowing us to discuss the presence of a nontrivial effective SOC inherent to the chirality. 
As we mentioned, a SOC that induces the mixture of spin-singlet and triplet Cooper pairs is essential for the EMChA in superconductivity. 
Through the model analysis, we experimentally validated an anomalously enhanced SOC with finite triplet mixing in our chiral superconductor, which points to the presence of sturdy spin-momentum locking in the Cooper pairs. 
In addition, we have confirmed 10$^3$ times enhancement in spin polarization during our previous CISS measurement\cite{nakajima2023} which also suggests stronger spin-momentum locking with respect to conventional Edelstein effect only from organic SOC\cite{he2020, edelstein1995}. 
Possible origins for the emergence of such gigantic spin-momentum locking in the present system without relying on the atomic SOC may include geometrical SOC that appears from the relativistic effect of a helical geometry\cite{shitade2020, hehl1990}, electronic exchange interactions in helical structures\cite{dianat2020}, coupling between vibration mode and electrons\cite{fransson2021} or the contribution of chiral phonons\cite{ishito2023}, all of which have been discussed in the context of effectively enhanced SOC in CISS effect. 
For any reasons, now one can clearly state that chiral superconducting systems possess larger spin polarization effects than other non-centrosymmetric systems in terms of SOC scalability.

In summary, the present study with the chiral organic superconductor has uncovered a gigantic EMChA around $T_{\rm{c}}$ whose amplitude surpasses previous records in inorganic polar systems; the chiral version of SDE below $T_{\rm{c}}$ whose amplitude is of similar magnitude of artificial polar superconductor comprising heavy elements; and a formation of two-critical-current superconductivity. 
All of these findings point to the appearance of triplet-mixed Cooper pairs correlating with strong chiral-type (namely, $\bm{k} \cdot \bm{\sigma}$-type) spin-momentum locking, which is very likely inherent to chiral symmetry rather than polar symmetry. 
The conclusion supports the giant spin polarization induced by triplet-mixed supercurrent, which can be also justified by the previously reported giant CISS effect. 
From a broader perspective, the present observations imply that a similar strong spin-momentum locking in molecular orbitals can be also explored to rationalize high spin polarization in molecular CISS effect. 
Note that electron orbitals in chiral molecules are coherent over the entire body which is similar to our superconductors. 
A definitive explanation of the mechanism behind the emergent SOC and mixed triplet pair requires further experimental and theoretical studies. 
Nonetheless, it is evident that these features are a striking source of efficient spin- and charge-rectification ability in the chiral systems. 
Emergence of such triplet-mixed Cooper pairs also pave the way to development of high critical field superconducting materials as well as novel type of quantum computers\cite{nayak2008}.

\section{ACKNOWLEDGMENTS}
We acknowledge valuable discussion with H. Kusunose. 
This work was supported by PRESTO from JST (Grant Number JPMJPR2356), Japan Society for the Promotion of Science (JSPS) (Grant Numbers 24K01331, 23H00291, 23H00091 and 21H01032), OML Project by the National Institutes of Natural Sciences (NINS program number, OML012301) and Special Project by Institute for Molecular Science (IMS program 23IMS1101).

%

\end{document}